# Atomically straight steps on vicinal Si (111) surfaces prepared by step-parallel current in the kink-up direction

S. Yoshida, T. Sekiguchi, and K. M. Itoh

*Department of Applied Physics and Physico-Informatics and CREST-JST, Keio University, 3-14-1, Hiyoshi, Kohoku-Ku, Yokohama 223-8522, Japan*

We demonstrate that annealing of a vicinal Si(111) surface at about 800 °C with a direct current in the direction that ascends the kinks enhances the formation of atomically straight step edges over micrometer lengths, while annealing with a current in the opposite direction does not. Every straight step edge has the same atomic configuration U(2,0), which is useful as a template for the formation of a variety of nanostructures. A phenomenological model based on electromigration of charged mobile atoms explains the observed current-polarity dependent behavior.

It has been shown that self-organized step structures on Si (111) vicinal substrates are ideal templates for the growth of a variety of nanostructures.[1] We recently proposed that atomic wires of $^{29}$Si isotopes with nuclear spins, which may be grown on such straight-step substrates,[2] can be utilized for practical quantum computation based on silicon.[3] Himpsel and co-workers were the first to develop such Si templates of highly ordered atomic step edges.[4,5] They applied dc current parallel to the steps to heat the vicinal Si(111) surface having a 1.1° miscut towards the $[\bar{1}\bar{1}2]$ direction and succeeded in preparing straight step edges with only one kink in a region extending over 8 μm in length.[4,5] Such kink removal was not observed in other studies where a DC current was applied in the direction perpendicular to the step edges,[6,7] confirming that a current direction parallel to the step edges plays a key role. On the other hand, following previously successful procedures,[4,5] which includes a specific miscut orientation and angle, a thermal sequence, and current direction parallel to step edges, we find the probability of success for obtaining perfectly straight, kink-free step edges to be about a half. In this letter, taking into account that there are two selectable directions, or polarities, parallel to the step edges for the dc current, as depicted in Fig. 1(a), we show that a so-called "kink-up" current results in a dramatic reduction of kinks over micrometer lengths while a "kink-down" current does not.

The vicinal Si(111) substrates for this study were prepared by polishing a commercially available flat (±0.1°) wafer of p-type Si to make a miscut of 1° towards a $[\bar{1}\bar{1}2]$ direction. In addition to the 1° polar miscut, an azimuthal misorientation of less than 3° from the $[\bar{1}\bar{1}2]$ towards either $[1\bar{1}0]$ or $[\bar{1}10]$ was intentionally introduced. Figure 1(b) shows an STM image of this vicinal surface after heating at 1260 °C by the kink-up current in ultrahigh vacuum. The step edges after such a cleaning procedure contain more than 30 μm$^{-1}$ kinks due to the azimuthal miscut and quenching from the highly mobile 1×1 phase. The same kink density has been obtained after 1260 °C cleaning with the kink-down current.

The first set of the thermal sequence consists of flash heating of the substrate up to 1260 °C followed by rapid cooling (within 3 s) to a temperature (typically 830 °C) below the 1×1 to 7×7 phase transition temperature ($T_c$ = 860 °C). A subsequent annealing at 830−800 °C for 10 h with a specific current direction as shown in Fig. 1(a) is performed. Finally, the sample is cooled down to room temperature in ~40 min. When the kink-up ([1$\bar{1}$0] in Fig. 1) current is applied, straight step edges are predominantly formed in the template area [Fig. 2(a)] at the expense of the kink aggregation (kink bunching) outside of this region [Fig. 2(a′)], which is about 1 μm away from the straight-step region [Fig. 2(a)] for the given in-plane misorientation ~3°. In contrast, the kink-down ([$\bar{1}$10] in Fig. 1) current results in a random distribution of small kinks [Fig. 2(b)] over millimeter lengths. In another set of thermal sequence, we first quench the sample from 1260 °C to room temperature to obtain the kink distribution shown in Fig. 1(b) and perform the 10-hour annealing at 830–800 °C to confirm the same current dependence of the kink redistribution as shown in Fig. 2. Thus we conclude that the critical factor for formation of straight step edges is annealing at 830–800 °C with the kink-up dc current for a sufficiently long time. The same kink-up annealing effect is confirmed for both azimuthal misorientations towards [1$\bar{1}$0] and [$\bar{1}$10].

Current-direction dependent step bunching phenomena have been reported previously when the current is passed perpendicular to step edges.[8-11] At temperatures sufficiently below $T_c$, a step-up dc current led to an alternating array of step bunches and broadened terraces, while a step-down current decomposed step bunches into rather uniformly distributed single steps.[10] In order to model our step-parallel current dependent kink behavior, we assume that adatoms during annealing are positively charged as reported for the same surface at different temperatures,[11,12] i.e., adatoms are to electromigrate in the direction of current due to the externally applied electric field. Moreover, the fact that every kink is terminated by the F half of the 7×7 unit cell as seen in Figs. 1 and 2 implies that the kink terminated by the U half is unstable on this vicinal surface. Therefore, we assume that emission of the F half atoms from the kink edge is quickly followed by the emission of the U half. Within this framework, we propose the mechanism for the straight-step broadening by the kink-up current and the kinks redistribution by the kink-down current as illustrated in Fig. 3. Atoms released from a kink drift in the direction of the current primarily along the step edge (solid arrow) due to the lower adsorption potential than that on a terrace. Figures 3(a) and 3(b) show a straight edge broadening mechanism with the kink-up current. Atoms released and migrating from the kinks labeled *A*, *B*, and *D* in (a) reconstruct the kinks at *B*, *C*, and *E*, respectively, in (b). However, atoms from the kink *C* in (a) has the larger probability of diffusing away from the step edge to the terrace (dashed arrow) because of the long traveling distance between *C* and *D*. Therefore atoms traveling from *C* may skip *D* and form a kink further above. Thus the step edge *C-D* in (a) is extended to *C-D'* (enlarged by 7*a*) as shown in (b), while the edge *D-E* is shortened to *D'-E* (by 7*a*). Of course atoms migrating larger distance and/or on narrower terraces have a large probability of descending steps. Such inter-step transport may be necessary for the formation of kink bunches shown in Fig. 2(a'). In this manner, the kink-up current enlarges preferentially the longer straight edges as seen in Fig. 2(a) at the expense of forming kink bunches in the very short straight edge regions as seen in Fig. 2(a'). The kink-down current annealing is just the opposite, as shown in Figs. 3(c) and 3(d). Atoms released and migrating from the kinks *F*, *G*, and *I* in (c) reconstruct the kinks *G*, *H*, and *J*, respectively, in (d), maintaining the interkink distances. However, atoms from *H* in (c) diffuse away from the step edge (dashed arrow) similarly to atoms from *C* in (a). This way, longer step edges [*H-I* in (c)] are shortened more efficiently [*H-I'* in (d)], thus leading to uniform distribution of kinks along the step edges as seen in Fig. 2(b).

The most important achievement of this paper for useful template formation is that our kink-up thermal procedure leads to an identical atomic structure at every straight step edge. Figure 4(a) shows a high-resolution STM image of this identical structure, which is schematically illustrated in Fig. 4(b). While the upper terrace is terminated with the boundaries (dimers) of the unfaulted halves of the 7×7 structure, the lower terrace has four adatoms per 7×7 unit-cell length at its corner. The shift of the 7×7 structure is $(2+2/3)b$ [$b/a = \sqrt{3}/2$, $a = 0.384$ nm] perpendicular to the step, but no shift is present parallel to the step. This atomic structure is referred to as U(2,0) according to Ref. 2.

In principle, the area of the atomically straight step region can be extended by reducing the azimuthal miscut. However, a sample we prepared recently with a very small but finite azimuthal miscut angle (<1°) contained kinks in both $[1\bar{1}0]$ and $[\bar{1}10]$ directions, making it impossible to define the kink-up current. The ability to polish the wafer precisely flat with a very small azimuthal miscut, by which one can achieve a small number of kinks distributed in the same direction, may be the key to fabricating a large template using our kink-up heating method.

In summary, we have developed a thermal procedure to reproducibly fabricate atomically straight $[\bar{1}\bar{1}2]$ steps on the vicinal Si(111) surface with an intentional azimuthal misorientation. Annealing at around 800 °C with a kink-up dc current formed the straight step edge regions much more effectively than a current in the kink-down direction. The atomically straight step edges have an identical atomic configuration, which is suitable as a template for the formation of a variety of nanostructure assemblies.

We would like to acknowledge H. Tochihara, S. Hasegawa, and I. Matsuda for fruitful discussions, and J. Lue and E. E. Haller for helpful discussions and reviewing. This work is supported in part by a Grant-in-Aid for Scientific Research in a Priority Area "Semiconductor Nanospintronics" Grant No. 14076215.

**Figure captions**

FIG. 1.  (a) A schematic of steps and kinks on the Si (111) vicinal surface and definition of the kink-up and kink-down current directions.  (b) The STM image of the as-cleaned vicinal surface with a 1° polar miscut towards the $[\bar{1}\bar{1}2]$ direction and the azimuthal miscut ~3° after 1260 °C flash cleaning followed by quenching to room temperature.  The area and the sample bias voltage are 580×580 nm$^2$ and +1.5 V, respectively.  The derivative of the topography is shown to emphasize the step edges.  Each dark line is a single step descending from left to right.

FIG. 2.  STM images after long annealing (~800 °C for 10 h) by dc current in the kink-up [(a), (a')] and kink-down [(b)] directions, respectively.  The kink-up current forms the straight-step region (a) at the expense of the kink-bunched region (a'), which is 1 μm away in step length.  The kink-down current dispersed the kinks uniformly as shown in (b) over millimeter lengths.  The derivative of the topography, recorded with the bias voltage of +1.5 V, is shown to emphasize the step edges.  Each dark line is a single step descending from left to right.

FIG. 3.  A schematic illustration of kink movements during annealing at ~800 °C, which explains the current direction dependence of the kink-redistribution observed in Fig. 2.  (a), (b) The straight edge broadening mechanism with the kink-up current.  (c), (d) The uniform distribution of kinks with the kink-down current.  Refer to the main text for details.

FIG. 4.  Atomic structure of the atomically straight step edge after the kink-up thermal procedure. (a) A high-resolution STM image of the 8×12 nm$^2$ area shows the arrangement of adatoms. (b) A schematic illustration of (a), which is referred to as U(2,0) (Ref. 2).  Adatoms of the upper and lower terraces are indicated by large and small circles, respectively.  The arrow indicates the displacement [$(2+2/3)b = 0.89$ nm] of the 7×7 unit cells across the step edge.

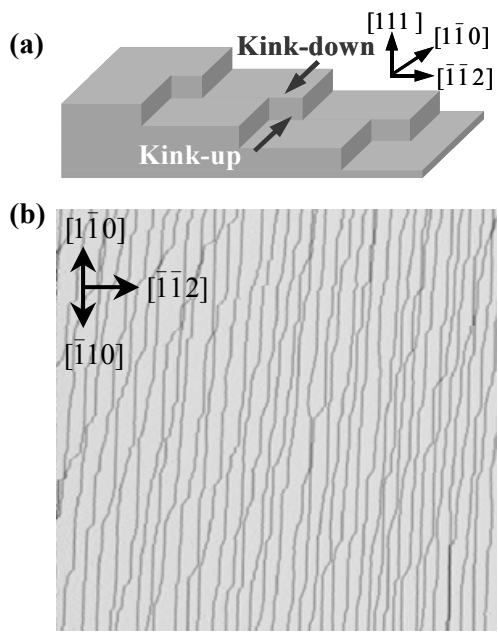

Fig. 1 Yoshida, Sekiguchi, and Itoh

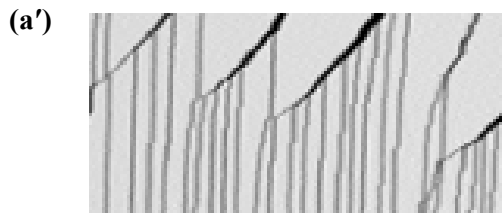

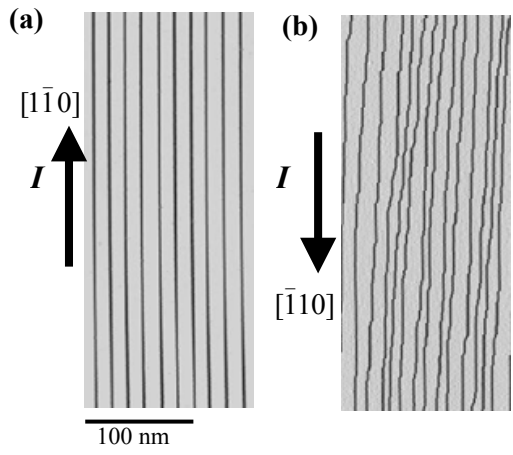

Fig. 2 Yoshida, Sekiguchi, and Itoh

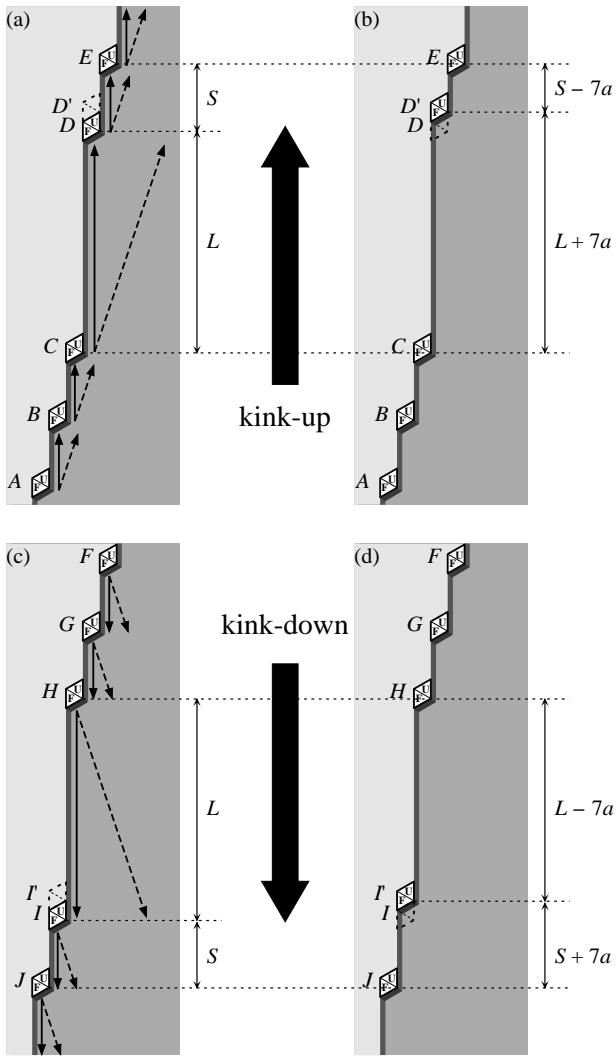

Fig. 3 Yoshida, Sekiguchi, and Itoh

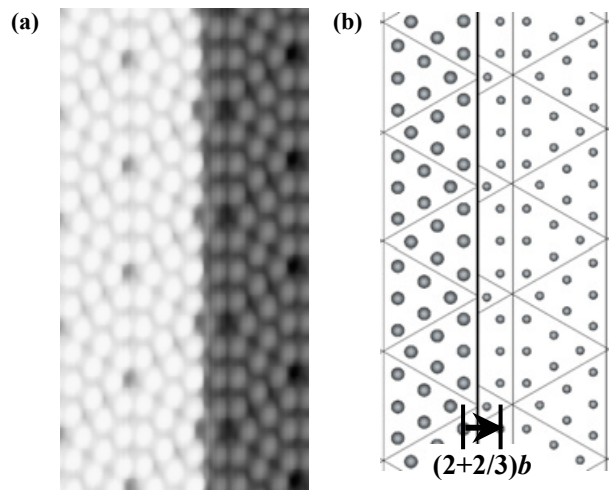

Fig. 4 Yoshida, Sekiguchi, and Itoh